\documentclass[a4paper]{jpconf}
\usepackage{graphicx}
\begin{document}
\title{Azimuthal correlations and collective effects in a heavy-ion collisions at the LHC energies}

\author{Ilya Selyuzhenkov}

\address{Research Division and ExtreMe Matter Institute EMMI,\\
GSI Helmholtzzentrum f\"ur Schwerionenforschung, Darmstadt, Germany}

\ead{ilya.selyuzhenkov@gmail.com}

\begin{abstract}
Recent highlights from the anisotropic flow and the azimuthal correlation measurements in a heavy-ion collisions at the LHC are presented.
Various flow harmonics measured for the charged and identified
particles versus transverse momentum,
pseudo-rapidity, and the collision centrality are reported.
New experimental probes of the local parity violation at the LHC energies
using the charge dependent azimuthal correlations are also discussed.

\end{abstract}

\section{Introduction}
An azimuthal anisotropic flow which describes collectivity among
particles produced in heavy-ion collision is recognized as
a key observable used to infer information about
the early time evolution of the nuclei interaction.
This FAIRNESS 2012 Conference proceedings highlight recent results
by the LHC experiments from the anisotropic flow
measurements in relativistic heavy-ion collisions at the TeV energy scale.
Experimental findings from the search for effects of
the local parity violation in strong interaction
using the charge dependent azimuthal correlations
with respect to the reaction plane at the LHC energies
are also discussed.

\section{Anisotropic flow fluctuations}
\begin{figure}[ht]
\begin{center}
  \includegraphics[width=.5\textwidth]{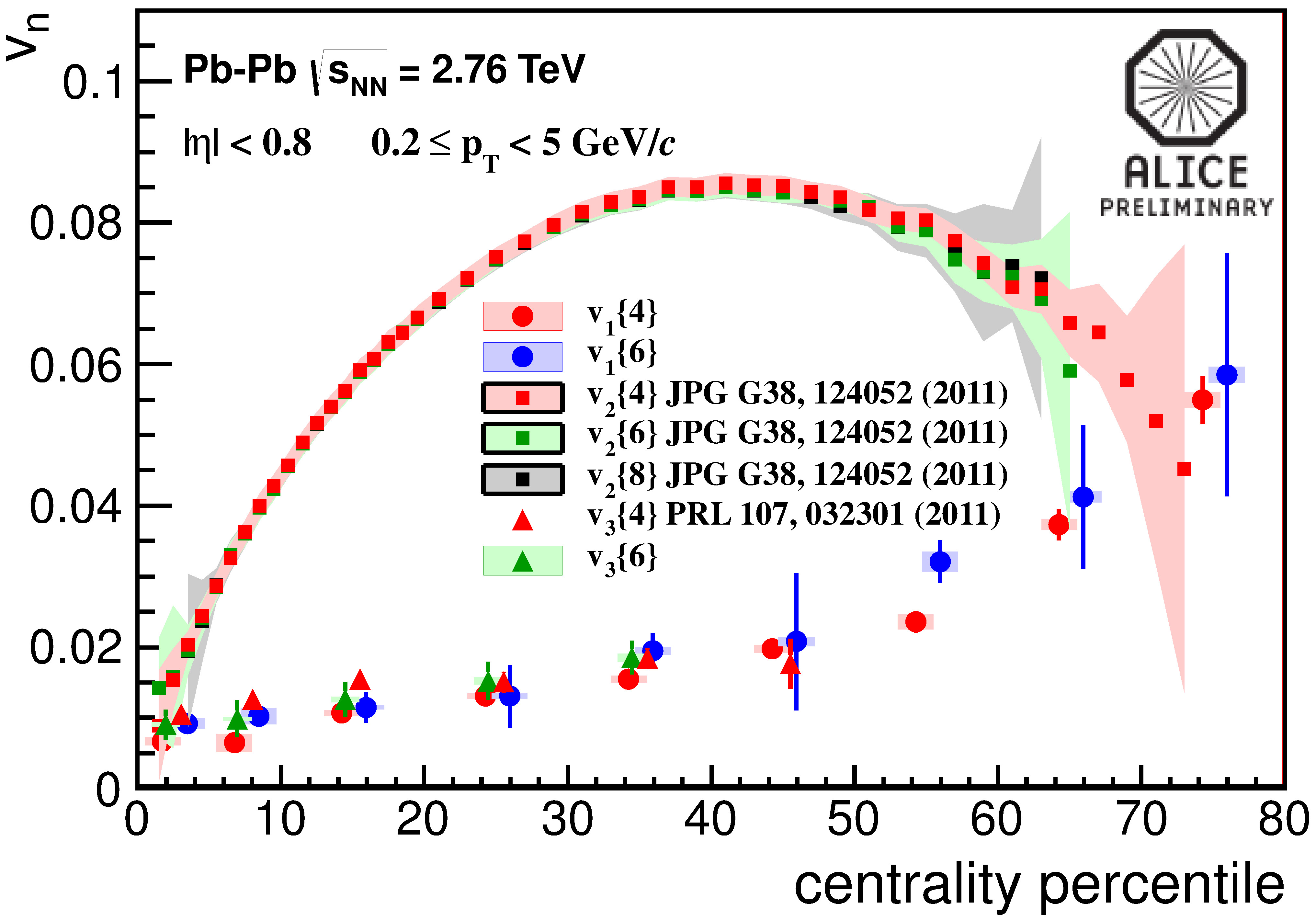}~%
  \includegraphics[width=.46\textwidth]{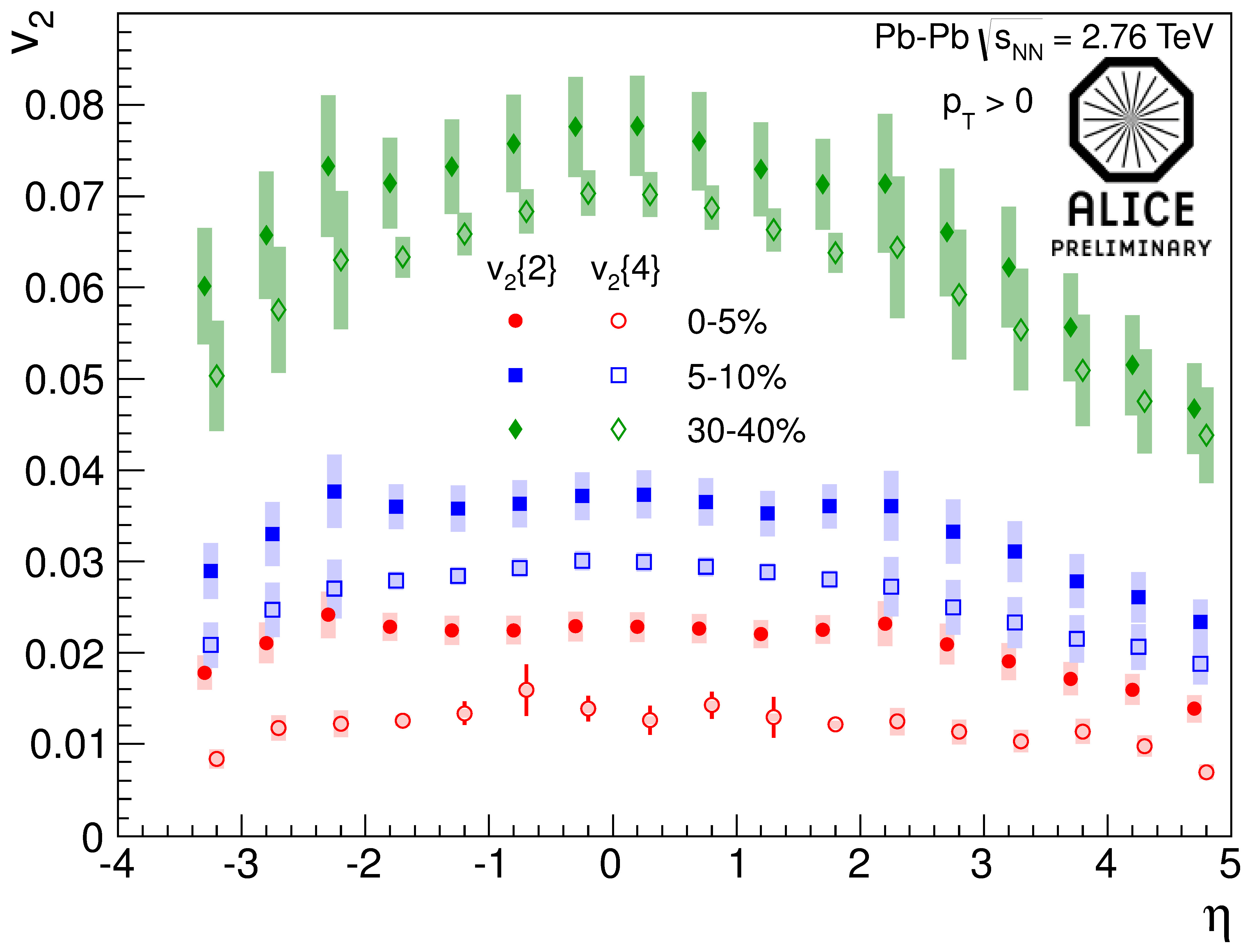}%
{\mbox{}\\\vspace{-0.2cm}
\hspace{1cm}\mbox{~} \bf (a)
\hspace{+6.7cm}\mbox{~} \bf (b)}
{\mbox{}\\\vspace{-0.1cm}}
\caption{Charged particle $v_1$, $v_2$, and $v_3$
measured by ALICE  \cite{Bilandzic:2012an,Hansen:2012ur} in Pb-Pb collisions at $\sqrt{s_{\rm NN}} = $~2.76~TeV.
(a) Multi-particle estimates of $v_1$, $v_2$, and $v_3$ at midrapidity vs. centrality.
(b) Two- and four-particle estimates of $v_2$ vs. pseudorapidity.
}
  \label{fig:1}
\end{center}
\end{figure}

It is commonly understood that the even-by-event
fluctuations in the initial energy density
of a heavy-ion collision plays an important role
in the development of the azimuthal asymmetries
in the momentum distribution of the produced particles.
Since fluctuations are mapped by the interaction among constituents
of the expanding medium into the final momentum space asymmetry,
the measurements of the anisotropic flow
provide a unique experimental information about
the properties of the created medium, its evolution,
and profile of the initial conditions in a heavy-ion collision.
Recently a number of new high statistics experimental results from LHC
which help to understand details of flow fluctuations become available.
Figure \ref{fig:1}(a) shows the results of the multi-particle correlation analysis
of the directed $v_1$, elliptic $v_2$, and triangular $v_3$, flow measured
for Pb-Pb collisions at  \mbox{$\sqrt{s_{\rm NN}}$ = 2.76~GeV}
by the ALICE Collaboration \cite{Bilandzic:2012an}.
Observed non-zero odd flow harmonics represent a significant effect of
the initial energy fluctuations especially for the most central collisions
where all harmonics become comparable to each other.
Agreement between the anisotropic flow estimates with 4, 6, and 8-particle correlations
put stringent constraints on the shape of the event-by-even flow
distribution.

\begin{figure}[ht]
\begin{center}
  \includegraphics[width=.47\textwidth]{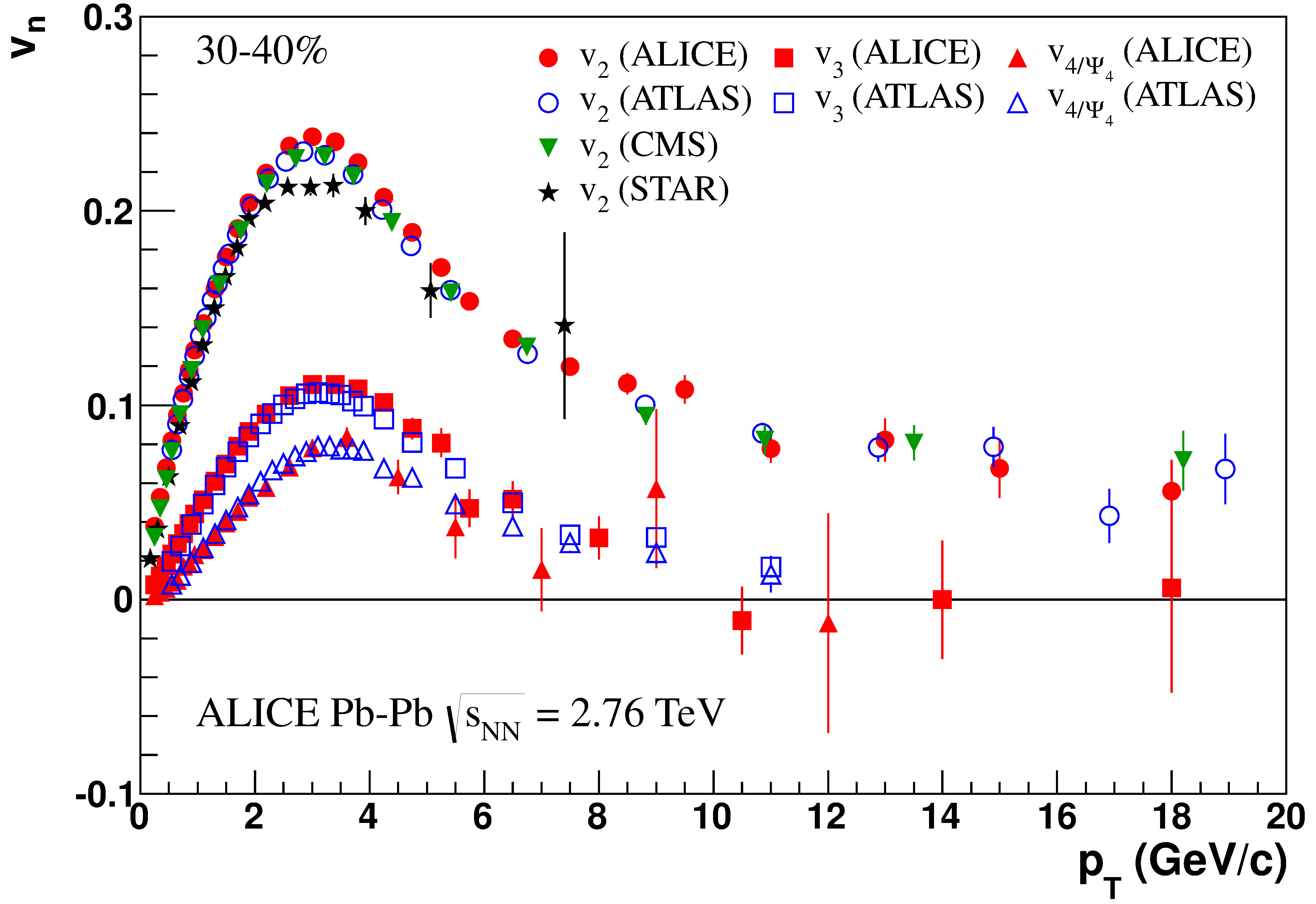}~%
  \includegraphics[width=.47\textwidth]{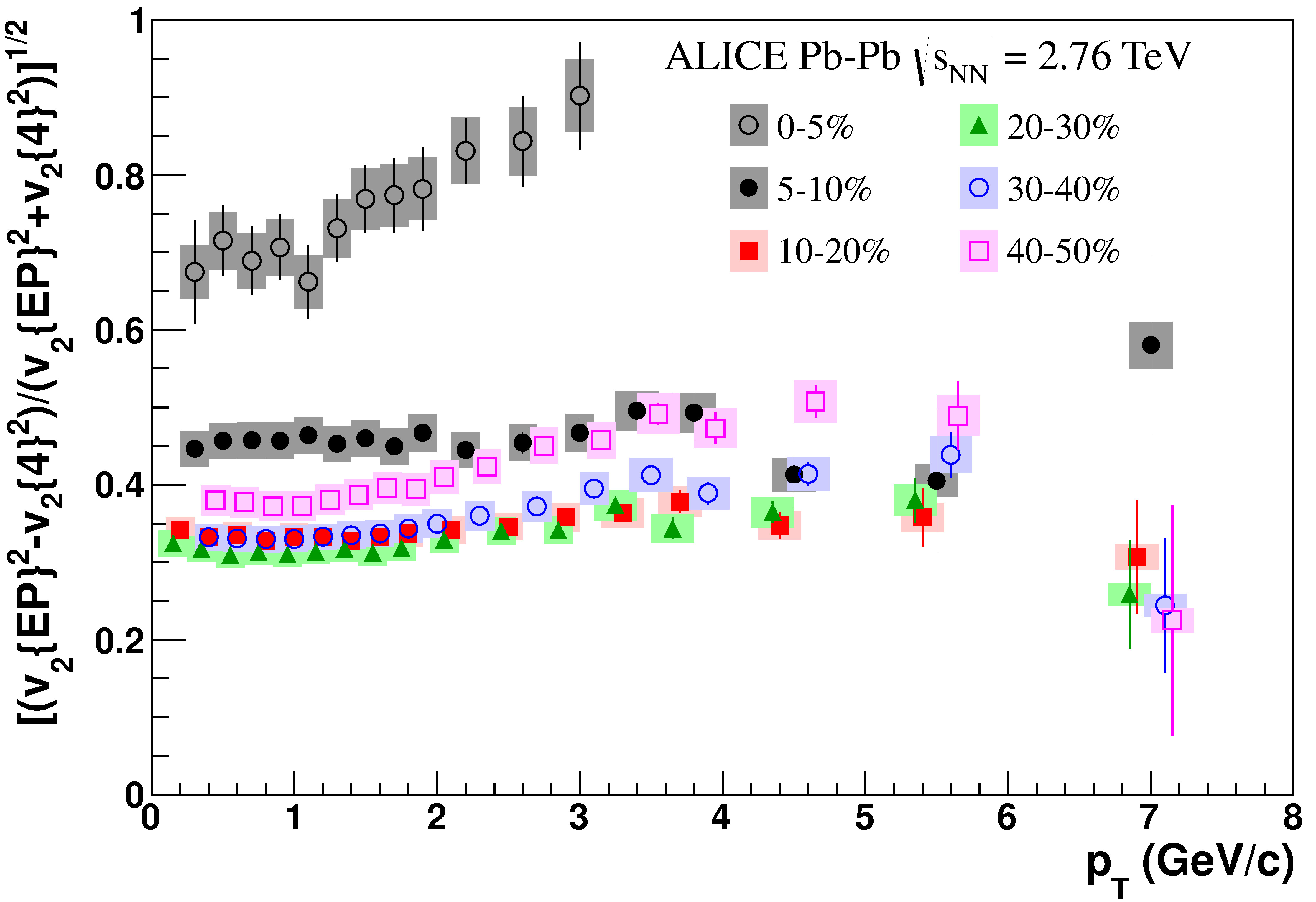}%
{\mbox{}\\\vspace{-0.2cm}
\hspace{1cm}\mbox{~} \bf (a)
\hspace{+6.7cm}\mbox{~} \bf (b)}
{\mbox{}\\\vspace{-0.1cm}}
  \caption{(a) Charged particle $v_2$, $v_3$, and $v_4$ measured for Pb-Pb collisions at $\sqrt{s_{\rm NN}} = $~2.76~TeV by the ATLAS \cite{ATLAS:2012at}, CMS \cite{Chatrchyan:2012ta}, and ALICE \cite{Abelev:2012di} Collaborations. (b) Estimate of the relative $v_2$ flow fluctuations as a function of the transverse momentum for different collision centrality classes.}
  \label{fig:2}
\end{center}
\end{figure}

Differential measurements of the anisotropic flow versus pseudorapidity $\eta$ and transverse momentum $p_{\rm T}$
provide additional information on the extend of flow fluctuations to forward rapidity and higher transverse momenta.
Such measurements performed by the ATLAS~\cite{ATLAS:2012at}, CMS~\cite{Chatrchyan:2012ta}, and ALICE~\cite{Abelev:2012di} Collaborations
are shown in Fig.~\ref{fig:1}(b) and Fig.~\ref{fig:2}.
The flow estimates obtained with the event plane, two and four-particle measurement techniques 
\cite{Chatrchyan:2012ta,Abelev:2012di,Hansen:2012ur}
suggest that relative flow fluctuations weakly depend on $p_{\rm T}$ and $\eta$.
Figure \ref{fig:2}(b) shows that fluctuations extend to higher $p_{\rm T}$ up to about 8~GeV/c,
while Fig. \ref{fig:1}(b) indicates a similar amount of relative fluctuations up to $|\eta|\sim 5$.

\begin{figure}[ht]
\begin{center}
  \includegraphics[width=.335\textwidth]{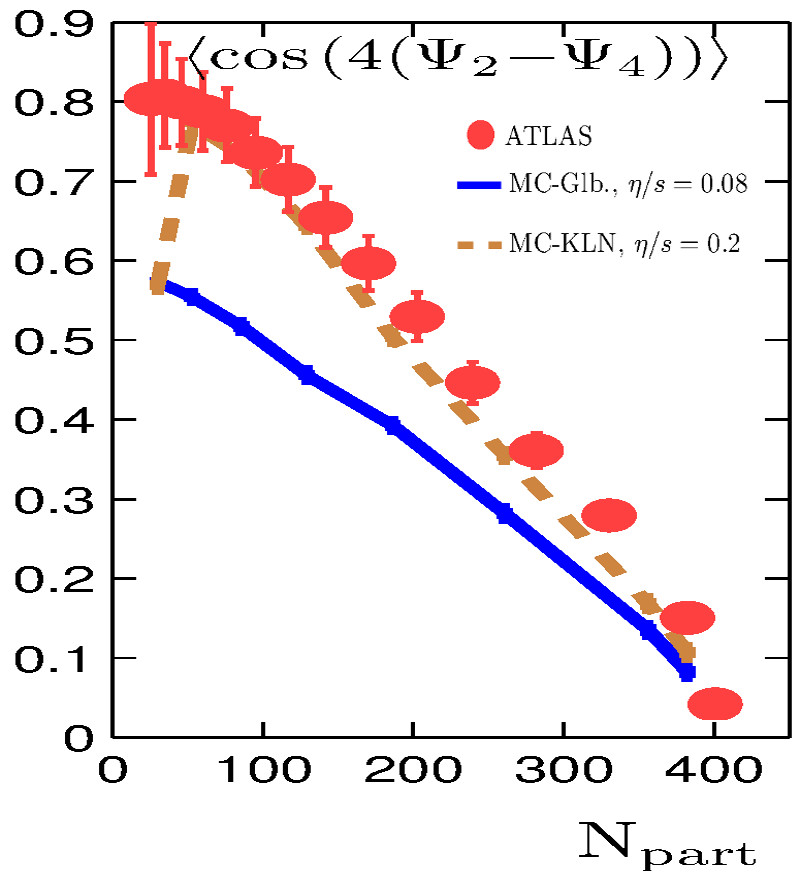}~~~%
  \includegraphics[width=.54\textwidth]{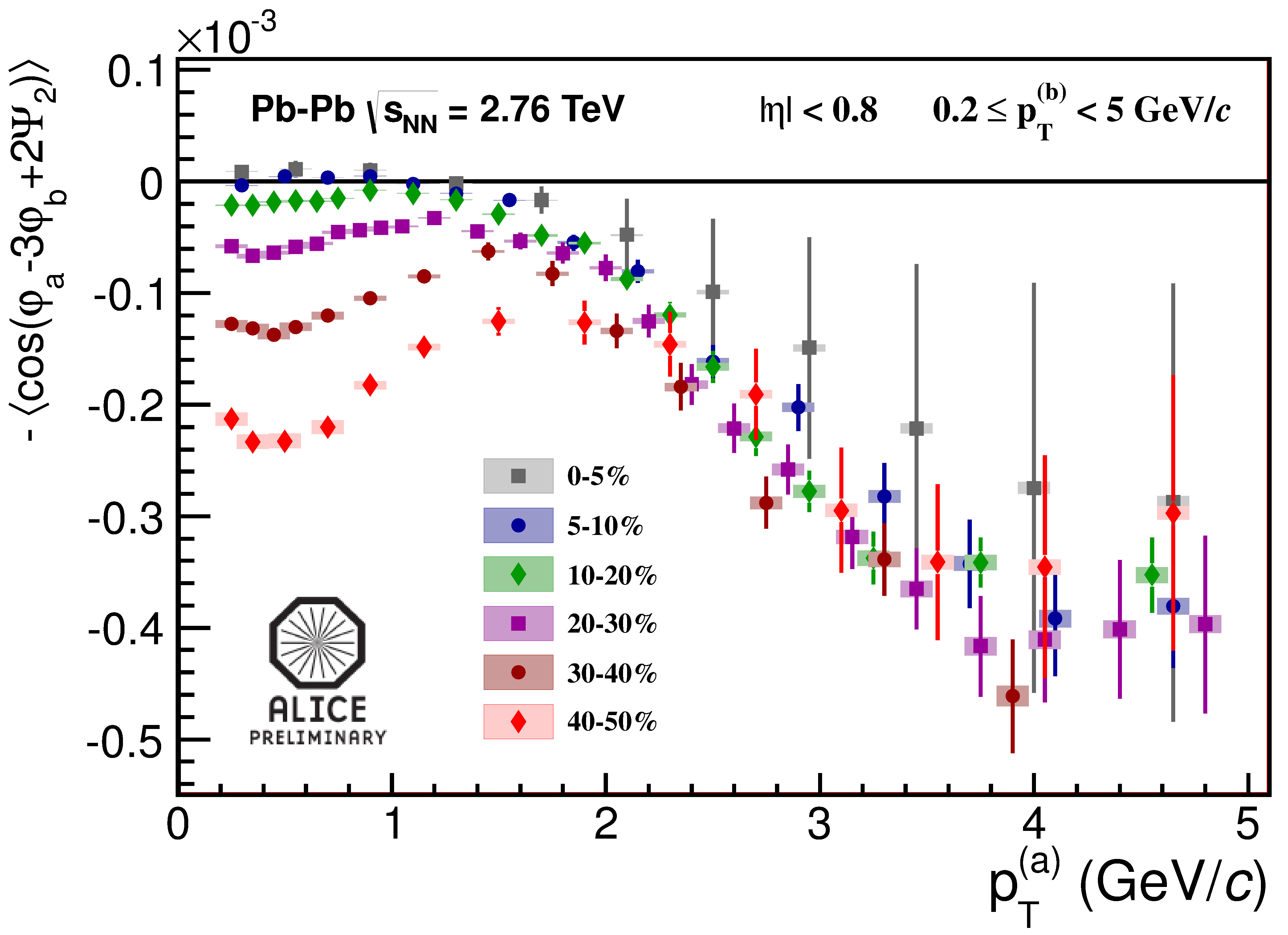}%
{\mbox{}\\\vspace{-0.4cm}
\hspace{-1cm}\mbox{~} \bf (a)
\hspace{+6.7cm}\mbox{~} \bf (b)}
{\mbox{}\\\vspace{-0.1cm}}
  \caption{(a)
2nd and 4th order event plane correlation vs. the number of participants (centrality)
measured by ATLAS \cite{Jia:2012sa} for
Pb-Pb collisions at $\sqrt{s_{\rm NN}} = $~2.76~TeV.
Results are compared with the event-by-event hydrodynamic model calculations.
(b) Three-particle azimuthal correlation proposed in \cite{Teaney:2010vd}
measured by ALICE \cite{Bilandzic:2012an} for the same collision system.
}
  \label{fig:3}
\end{center}
\end{figure}

Further constraints on the effects of the initial energy fluctuations is provided by the measurements
of the multi-particle mixed harmonic correlations and the correlation between the different order symmetry planes.
Figure~\ref{fig:3} shows corresponding results by the ATLAS~\cite{Jia:2012sa} and ALICE~\cite{Bilandzic:2012an} Collaborations in comparison with the expectations from the event-by-event hydrodynamic model calculations using
two different profiles of the initial condition~\cite{Qiu:2012uy}.
Observed correlations between the two and three planes of symmetry
and their qualitative agreement with the  hydrodynamic model calculations
points on their strong sensitivity to the details of the fluctuating initial energy profile.
Among other promising developments in the fluctuations studies are the shape measurements
of the event-by-event flow distributions (see e.g. \cite{Jia:2012ve}), and
the measurements of physics observables for event classes selected based on their azimuthal shapes.
This new type of study suggested in \cite{Schukraft:2012ah} was successfully applied in the real data analysis
by the ALICE Collaboration \cite{Dobrin:2012zx,Milano:2012qm}.

\section{Elliptic flow of identified particles}
Viscous hydrodynamics is considered to be a relevant model to describe a thermalized phase
in the time evolution of the system created in a heavy-ion collision.
\begin{figure}[ht]
\begin{center}
  \includegraphics[width=.44\textwidth]{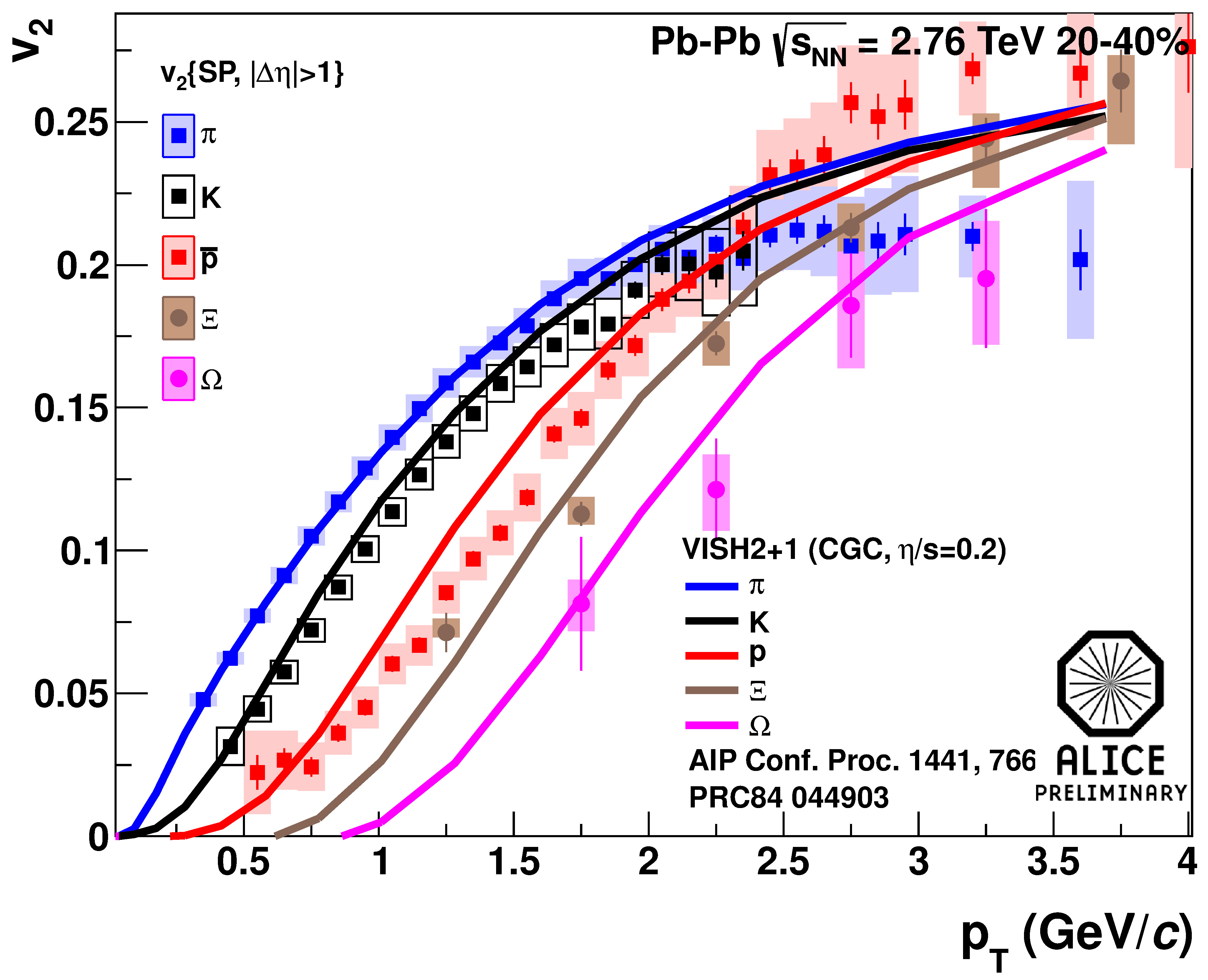}\mbox{~~}%
  \includegraphics[width=.43\textwidth]{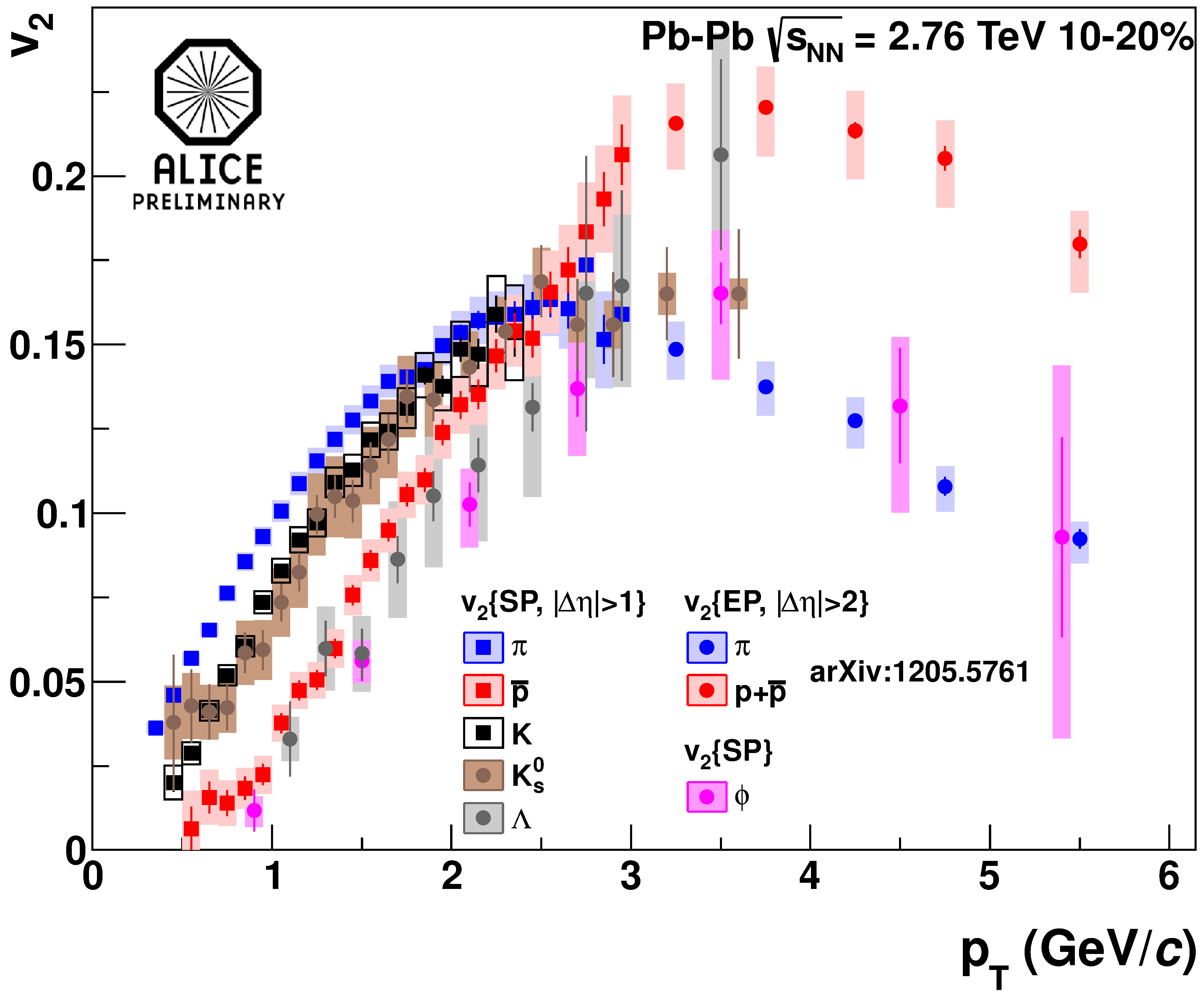}%
{\mbox{}\\\vspace{-0.4cm}
\hspace{1cm}\mbox{~} \bf (a)
\hspace{+6.7cm}\mbox{~} \bf (b)}
{\mbox{}\\\vspace{-0.1cm}}
  \caption{Identified particle $v_2(p_{\rm T})$ measured by ALICE \cite{Noferini:2012ps} for Pb-Pb collisions at $\sqrt{s_{\rm NN}} = $~2.76~TeV.
Charged pion, kaon, and proton $v_2$ compared to that of (a) $\Xi$ and $\Omega$ and the hydrodynamic model calculations \cite{Shen:2011eg}, and (b) with $v_2$ of $\phi$-meson, $K^0_s$ and $\Lambda$/$\bar\Lambda$ particles.}
  \label{fig:4}
\end{center}
\end{figure}
Applicability of the hydrodynamic description at the LHC energies
can be tested by measuring the particle mass dependence of the elliptic
and triangular flow at small transverse momenta, $p_{\rm T}<2-3$~GeV/$c$.
Figure~\ref{fig:4} shows $v_2(p_{\rm T})$ of pions and anti-protons
in comparison to that of strange
($K^0_s$, $\Lambda$/$\bar\Lambda$) and multi-strange ($\phi$, $\Xi$, and $\Omega$) particles.
The main trends of the observed mass splitting of $v_2$ at low transverse momenta is
reproduced by viscous hydrodynamic model calculations \cite{Shen:2011eg}
with a color glass condensate initial condition (see Fig.~\ref{fig:4}(a)).
The flow of heavier particles (proton, $\phi$-meson, $\Xi$ and $\Omega$)
is more sensitive to the hadronic rescattering phase
and the agreement with data improves when adding
this additional phase into the model calculations (see Fig. 2 in \cite{Noferini:2012zz}).
$v_2$ of $\phi$-meson which is shown in Fig.~\ref{fig:4}(b)
follows the mass splitting expected from the hydrodynamics at lower transverse momenta,
while its magnitude is close to $v_2$ of other mesons (pions)
in the intermediate region of $p_{\rm T} \sim 3-6$~GeV/$c$.
Such behavior suggest the constituent quark number ($n_q$) scaling of $v_2$
which is expected for mesons and baryons produced via quark coalescence
in a phase of the deconfined quarks and gluons. This scaling
may indeed holds at the LHC within 20\% at the $p_{\rm T}/n_q \sim 1.2$~GeV/$c$
(see Figs.~(4) and (5) in~\cite{Noferini:2012ps}).

\section{Probes of local parity violation in strong interaction}
The parity symmetry violation by the strong interactions
remains one of the open fundamental questions about the quantum chromodynamics.
It is argued \cite{Kharzeev:2007jp} that the parity
symmetry can be locally violated in a heavy-ion collisions
what will result in the experimentally observable separation of charges
along the extreme magnetic field generated by the moving ions,
the so called chiral magnetic effect (CME).
As an experimental probe of the CME it was proposed \cite{Voloshin:2004vk} to
use the azimuthal correlations with respect to the collision reaction plane
which is perpendicular to the magnetic field generated in the collision.
The STAR~\cite{Abelev:2009ac} at RHIC and the ALICE~\cite{Abelev:2012pa} at the LHC observed
a clear charge dependence of the two-particle correlation with respect to the reaction plane.
The measured observable is parity even and thus is sensitive to effect unrelated to the parity symmetry violation.
% The case of $m=-2$ corresponds to the correlator $\langle \cos(\phi_{\alpha}+\phi_\beta-2\Psi_{2})\rangle$ which was originally proposed in \cite{Voloshin:2004vk}.
%
\begin{figure}[ht]
\begin{center}
  \includegraphics[width=.53\textwidth]{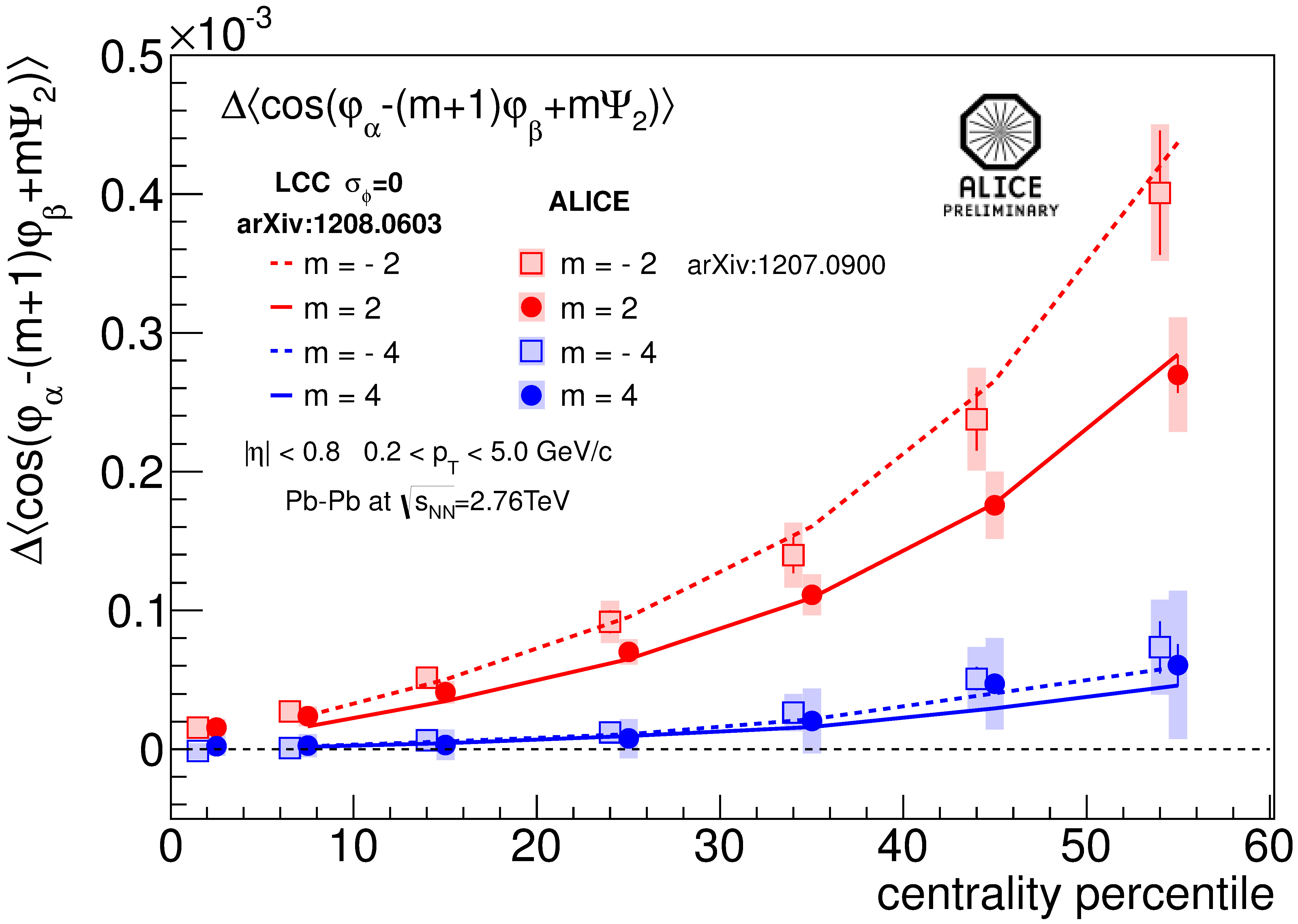}%
  \includegraphics[width=.38\textwidth]{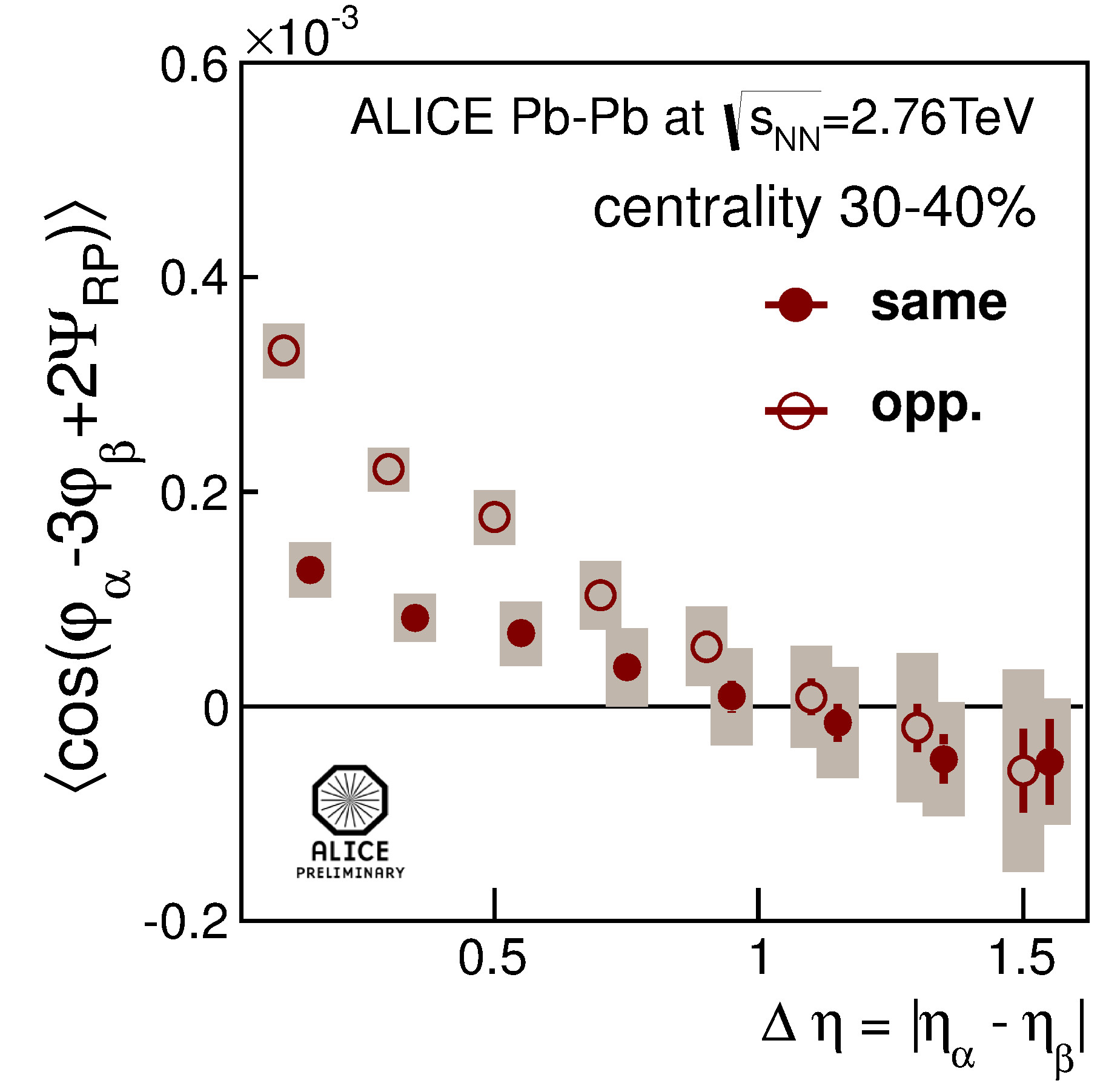}%
{\mbox{}\\\vspace{-0.3cm}
\hspace{2cm}\mbox{~} \bf (a)
\hspace{+6.7cm}\mbox{~} \bf (b)}
{\mbox{}\\\vspace{-0.1cm}}
  \caption{
Charge dependent correlations measured by ALICE~\cite{Hori:2012hi} for Pb-Pb collisions at $\sqrt{s_{\rm NN}} = $~2.76~TeV.
(a) The same minus the opposite charge correlations compared to the local charge conservation model \cite{Hori:2012kp}.
(b)~Correlation $\langle \cos(\phi_{\alpha}-3\phi_\beta+2\Psi_{2})\rangle$
vs. rapidity separation~\cite{Hori:2012hi}.
}
  \label{fig:5}
\end{center}
\end{figure}
Recently the ALICE Collaboration extended \cite{Hori:2012hi} the measurements
with the new set of the mixed harmonic correlators
$\langle \cos(\phi_{\alpha}-(m+1)\phi_\beta+m\Psi_{m})\rangle$ (for $m=\pm 2, \pm 4$)
and the double harmonic correlator $\langle \cos (2\phi_{\alpha}+2\phi_\beta-4\Psi_{4})\rangle$,
where $\phi_{\alpha,\beta}$ is the azimuthal angle, $\alpha,\beta$ is the charge of the particle, and $\Psi_{m}$ is the $m$-th order collision symmetry plane angle.
Figure~\ref{fig:5}(a) shows the correlation difference $\Delta\langle \cos(\phi_{\alpha}-(m+1)\phi_\beta+m\Psi_{m})\rangle$
between the same and the opposite charge combinations.
The expected CME contribution to the correlator $\langle \cos(\phi_{\alpha}+\phi_\beta-2\Psi_{2})\rangle$ ($m=-2$ in Fig.~\ref{fig:5}) at the LHC energies is based on the extrapolation of the RHIC results.
Depends on the model assumptions, it varies from about 20\% \cite{Toneev:2010xt}
up to the full strength \cite{Kharzeev:2007jp,Zhitnitsky:2010zx,Zhitnitsky:2012im} of the correlations
observed at the LHC if one assumes that magnetic field  flux is independent of the collision energy. 
It is noted \cite{Pratt:2010gy} that effects of local charge conservation
modulated by the anisotropic flow can be responsible
for a significant part of the measured correlations.
The statement is also supported by model calculations \cite{Hori:2012kp} shown in Fig.~\ref{fig:5}(a).
The charge independent part of the correlations can be dominated
by effects of flow fluctuations~\cite{Teaney:2010vd}, although the
differential study of the charge dependent correlation
shown in Fig.~\ref{fig:5}(b) reveals strong dependence
on the pair separation in pseudorapidity which is not typical for flow fluctuations.
At the moment none of the models is able to reproduce simultaneously the charge
dependence and the charge insensitive baseline of the measured correlations.

\section{Summary and outlook}
The anisotropic flow and the multi-particle azimuthal correlation measurements at the LHC
help to clarify the role of the initial energy fluctuations in
the multiple particle production in a heavy-ion collision.
The  particle type and the mass dependence of the anisotropic flow suggest that
the system produced in a heavy-ion collision
is strongly coupled and evolved via the deconfined phase of quarks and gluons.
The new measurements of the mixed harmonic charge dependent azimuthal correlations helps
to constrain possible background sources in the search for
effects of the parity symmetry violation in the QCD.

\section*{Acknowledgements}
Supported by the Helmholtz Alliance
Program of the Helmholtz Association, contract HA216/EMMI ``Extremes of
Density and Temperature: Cosmic Matter in the Laboratory".

\end{document}